# Astro2020 APC White Paper

# Robotic laser adaptive optics for rapid visible/near-infrared AO imaging and boosted-sensitivity low-resolution NIR integral field spectroscopy

**Corresponding Author:**
Christoph Baranec (University of Hawaii; baranec@hawaii.edu; 808-932-2318) on behalf of the Robo-AO, Robo-AO-2 and Rapid Transient Surveyor Teams (http://www.ifa.hawaii.edu/Robo-AO/team.html)

**Additional Information:**
http://robo-ao.org

**Introduction:** Large area surveys will dominate the next decade of astronomy, and the main limitation to science will be the thorough followup and characterization of their extremely numerous discoveries. The deployment of robotic laser adaptive optics on our nation's mid-sized telescopes will be crucial for the sensitive and rapid characterization of these survey targets. The soon to-be-commissioned Robo-AO-2 system on Maunakea (Baranec et al. 2018) will combine near-HST resolution imaging across visible and near infrared wavelengths and will enable high-acuity, high-sensitivity follow-up observations of several tens of thousands of objects per year. Robo-AO-2 will also respond to target-of-opportunity events within minutes, minimizing the time between discovery and characterization, and will interleave different programs with its intelligent queue. The Robo-AO-2 imaging system can be replicated for <USD$2M, making it a cost effective way of upgrading the capabilities of modest aperture telescopes.

An often overlooked benefit of AO correction is the increased concentration of energy in wavelengths where sky-brightness dominates. E.g., AO correction will boost the infrared point-source sensitivity of a spectrograph on the UH 2.2-m telescope by a factor of seven for faint targets, giving it the H-band sensitivity of a 5.7-m telescope without AO, making the classification of transient objects much more efficient. This will also serve as a pathfinder for a future implementation on 8-30 m class telescopes - crucial in the era of LSST where larger apertures will be necessary to fully characterize the wealth of new discoveries (Kulkarni 2002).

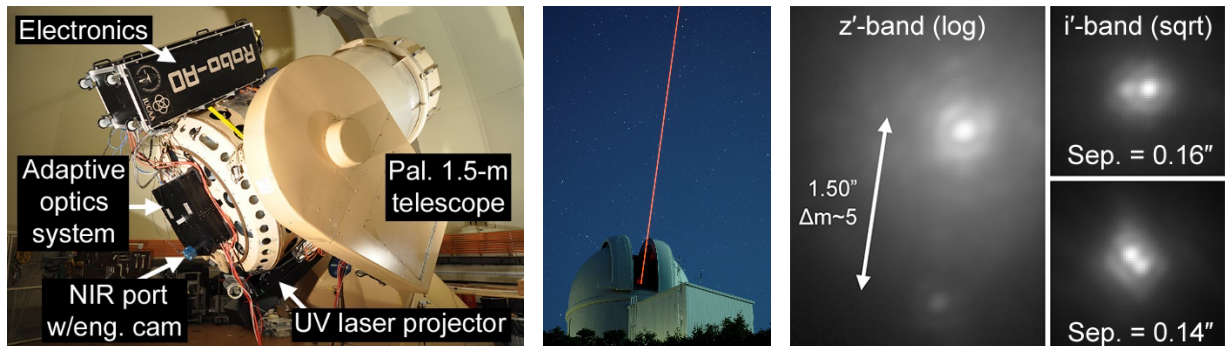

**Figure 1. Left:** The prototype Robo-AO system on the 1.5-m telescope at Palomar. **Center:** The UV laser propagating on sky as seen by a modified DSLR camera. **Right:** Robo-AO corrected visible light images of binary stars.

**Technology Heritage:** Our team first commissioned Robo-AO at the 1.5-m Palomar telescope in 2011, and then achieved fully robotic operation in mid-2012, having the system perform 120 observations in a row with no human intervention (Figure 1; Baranec et al 2013, 2014, 2017). Since then, additional optimizations have reduced observing overheads to approximately 20 s - compared to large telescope laser AO system which have typical overheads of 5-20 minutes - thus enabling the observation of 200-250 targets in a night with no weather losses. Baranec et al. (2015) also went on to demonstrate the first use of infrared avalanche-photodiode arrays for imaging and wavefront sensing with Robo-AO, which are now being implemented in other adaptive optics and interferometric systems (e.g., Keck IR pyramid wavefront sensor, SCExAO at Subaru, VLT-GRAVITY). Baranec and Riddle were awarded three U.S. patents, nos. 9,279,977, 9,563,053, and 9,405,115, for Robo-AO technologies. And according to the NSF, "Robo-AO has the distinction of being the most widely acknowledged NSF Advanced

Technologies and Instrumentation award among its more than 30 year history." (Kurczynski & Neff 2018 & Kurczynski priv. comm.).

**Technology Development:** The team moved Robo-AO to the 2.1-m telescope at Kitt Peak in 2015 for a 2.5 year deployment (Jensem-Clem et al. 2018), and it is being commissioned now on the Univ. of Hawaii 2.2-m telescope at Maunakea. Other adaptive optics systems have been directly based on Robo-AO or its technologies: the KAPAO natural guide star adaptive optics system, built primarily by undergraduate students at Pomona College (Severson et al., 2013); the iRobo-AO system being commissioned at the 2-m IGO telescope in India (https://instru.iucaa.in/index.php/igo/12-igo/28-iroboao); plans for the SOAR Robotic Adaptive Optics system (Ziegler et al. 2016, Law et al. 2016); and a version of Robo-AO-2 being built for the US Naval Observatory at Flagstaff, AZ. Additionally, the core Robo-AO software is now used for the Kitt Peak EMCCD demonstrator (Coughlin et al. 2019) and for the Zwicky Transient Facility (Bellm et al. 2019).

**Past Scientific Accomplishments:** Data acquired with Robo-AO is used in 40 refereed scientific papers, that represent adaptive optics images of approximately 12,500 unique targets (see http://www.ifa.hawaii.edu/ Robo-AO/pub_astro.html) and include several of the largest adaptive optics surveys to date. This includes the Robo-AO *Kepler* Candidate survey that is the only survey that has searched all the *Kepler* planet candidates for nearby stars. Other high-resolution surveys (e.g., Adams et al. 2013, Lillo-Box et al. 2014, Horch et al. 2014, Kraus et al. 2016, Furlan et al. 2017 and many more), have been targeted at deeper observations of much smaller number of targets, but the majority of the *Kepler* planet candidates have only been observed by Robo-AO. The Robo-AO team has published six refereed papers on the main survey results (Law et al. 2014, Baranec et al. 2016, Atkinson et al. 2017, Ziegler et al. 2017, 2018a, 2018b), two papers on selected subsets of the *Kepler* stars (Schonhut-Stasik et al. 2017, 2019), and contributed to a further fifteen refereed papers on transiting planets discovered by *Kepler*. The Robo-AO survey is the largest-ever AO survey, by far, covering a factor of seven more targets than the largest previous surveys.

The legacy of the Robo-AO *Kepler* Candidate Survey high-resolution survey is just starting, with the majority of the nearly 200 citations using the presence or absence of blended stellar companions detected by Robo-AO that are responsible for underestimates of planetary radii or that are false-positives. To increase the accessibility of the survey results, data were loaded onto the Community Follow up Program, https://cfop.ipac.caltech.edu/, and to the public website at http://roboaokepler.org/. Recent highlights include empirically validating the sensitivity of Gaia-DR2 to close separation binaries (Ziegler et al. 2018c) and to providing the most accurate planetary radius correction factor for KOI-4, in visible light where the *Kepler* light curves are measured (Chontos et al. 2019; famously known as *Kepler's* first exoplanet detection).

**Key Future Science Goals and Objectives:** Robo-AO-2 will enable a broad spectrum of unique and previously infeasible science with its combination of acuity, efficiency and flexible scheduling (Table 1). For more detail on the science cases, see Baranec et al. 2018.

| Science Topic | Wavebands | Number and/or Cadence | Investigators |
|---|---|---|---|
| IR transient characterization: e.g., novae, WRs, BDs, EM-GW | g' - H | Rapid response, declining follow-up cadence | Kasliwal, Moore, Lau |
| Wide exoplanets and brown dwarfs | H | 5,000-10,000 targets | Liu |
| Transit exoplanet hosts (e.g., TESS) | Broadband vis. | >10,000 targets | Baranec, Law |
| Discovering/monitoring L. Quasars | i' and H | >25,000 + monitoring 3 ngt./mo. | Griffiths, Ofek |
| Asteroseismology and multiplicity | Broadband vis. | Several thousands | Huber, Baranec |
| Planetary monitoring | g' - z' | Snapshot 2-3 times/night | dePater, Hammel, Simon, Wong |
| Stellar Multiplicity from large surveys | Discovery: r', i' Colors: g' - H | >1,000 (PS1) >10,000 (Gaia) | Magnier (w/PanSTARRS) Law (w/Gaia) |
| Multiplicity in stellar clusters | z' – H | Several hundred / cluster | Connelley, Reipurth |
| Monitoring Jets/Outflows/Shocks | Narrowband | Several times per year | Reipurth, Hodapp |
| Vetting archival surveys for blends | r', i', or J | 300 per year | Kuchner (w/DiskDetective) |
| Astrometric microlensing | Y, J, or H | Dozens of high cadence event/yr | Dekany (w/ZTF) |
| Small body nucleus characterization, exopause searches, surface minerology | g' - H | Few night response, ~10 Manx, Centaurs and comets per year | Meech |

**Table 1.** A sampling of the breadth of science to be enabled by the new Robo-AO-2 system.

**Technical Overview:** The Robo-AO-2 system will combine near-HST resolution across visible and near-infrared (NIR) wavelengths ($\lambda$ = 400 – 1800 nm), unmatched observing efficiency, and extensive, dedicated time on the UH 2.2-m. It will enable high-acuity, high-sensitivity follow-up observations of several tens of thousands of objects per year. Robo-AO-2 will also respond to target-of-opportunity events within minutes, minimizing the time between discovery and characterization, and will interleave different programs with its intelligent queue.

The design of the Robo-AO-2 system is based on the successful prototype and makes several improvements to the performance and capability. Robo-AO-2 will comprise a UV laser projector, a Cassegrain mounted adaptive optics system with low-noise, high-speed visible and infrared imaging arrays that double as tip-tilt sensors (see Table 2), a new reconfigurable stellar wavefront sensor, and a set of electronics and control computer with additional functionality.

| Detector | Format | Field | Pix.Scale | ReadNoise | Full frame rate | Tip-tilt rate | Initial filters |
|---|---|---|---|---|---|---|---|
| EMCCD | $1024^2$ | 31″×31″ | 0.030″ | <1e- | up to 26 Hz | to 500 Hz | g', r', i', z', H$\alpha$, SII, OIII, LP600 |
| SAPHIRA | 320×256 | 17″×14″ | 0.055″ | <1e- | up to 400 Hz | to 8 kHz | Y, J, H, FeII, Y+J+H |

**Table 2.** The Robo-AO-2 imaging/tip-tilt detectors will enable high acuity imaging over $\lambda$=400-1780nm.

**Technology Drivers:** In addition to the forthcoming imaging system, we have designed the Rapid Transient Surveyor (RTS) spectrograph to work with Robo-AO-2 that is optimized for quick target acquisition, high throughput, and streamlined calibration and data reduction (Figure 3; Baranec et al. 2016b; spectrograph leads: S. Wright, S. Chen). The spectrograph makes use of an Integral Field Unit (IFU) that divides the traditional image plane into spatial pixels, "spaxels", where each spaxel is dispersed to generate an individual spectrum on the detector. The RTS integral field spectrograph (IFS) is designed to have an average spectral resolution of R~100 that simultaneously covers the near-infrared band pass from 840 to 1830 nm with a spatial sampling of 0.15″ per spaxel. The primary components of an IFS are: relay optics; a method that samples the image plane (e.g., lenslet array, slicer, fibers); dispersing element; collimator and camera optics; and detector. Our team has developed a design for the RTS IFS that maximizes the use of the pixels of a Hawaii-2RG (18μm pixel size, 2048x2048) detector, while also allowing for the maximum field of view with a very conservative separation between neighboring spectra to allow for high-fidelity data reduction. The Hawaii-2RG will operate in a separate cryostat from GL Scientific, with a single band pass filter mounted in front of the detector, and use a PB-1 array controller and associated software (developed and built at UH). The IFS is designed to have no moving opto-mechanical mechanisms (e.g., filter or grating stages) to reduce the overall complexity and cost of the instrument, and to allow for stable spectra.

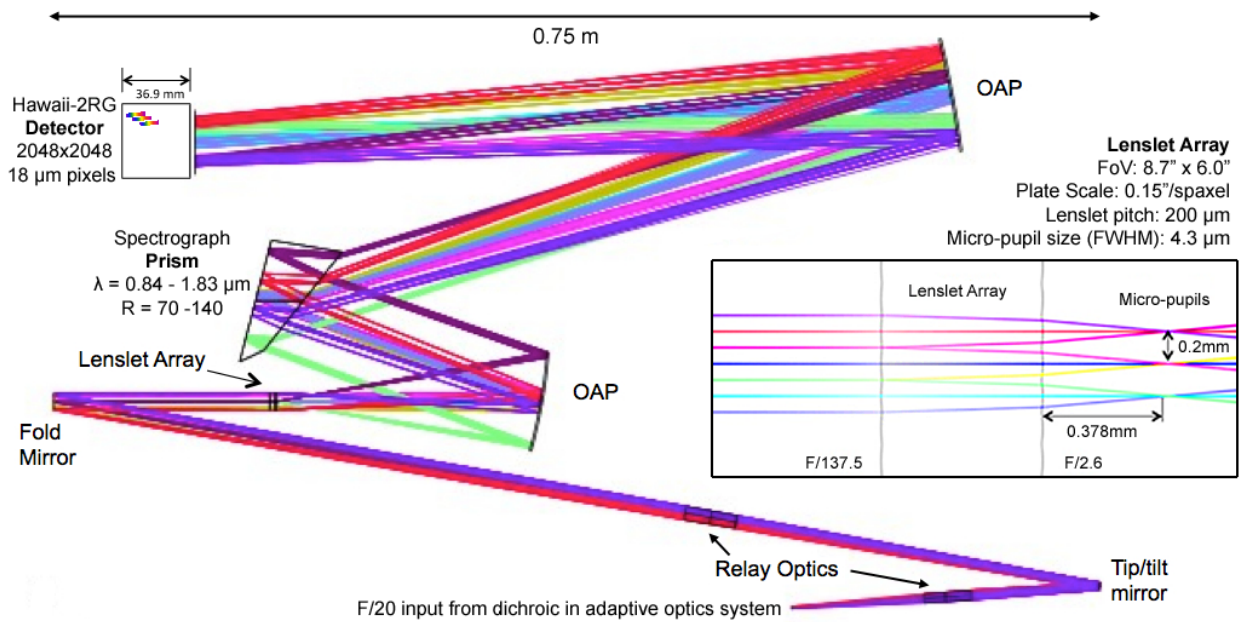

**Figure 3.** Optical diagram of the RTS integral field spectrograph illustrating the primary optical components. The F/20 input from the AO system enters at the bottom and the tip-tilt mirror is placed at the intermediate pupil of the F/137.5 relay optics that generates the 0.15″ plate scale at the location of the 58×40 lenslet array. **(Inset):** The ray trace of 3 adjacent lenslets. The F/137.5 beam enters from the left and enters the left surface of the lenslet that has a thickness of 1mm and the right surface of the lenslet has the majority of the power on the backside to concentrate the light down to well separated pupil images past the array to f/2.6 outgoing beam. Each lenslet produces a micro-pupil that feeds a standard spectrograph that has a single fixed prism (R~70-140) that disperses the near-

infrared bandpass (840 - 1830 nm) to 2,320 individual spectra on to a Hawaii-2RG detector.

The expected signal-to-noise (SNR) of RTS is greatly enhanced when paired with the Robo-AO-2 system. This is primarily due to the diffraction-limited PSF that the AO system delivers, which allows for smaller extraction apertures and thus smaller backgrounds. A comparison of the expected SNR for both the seeing-limited and AO-fed case is shown in Figure 4 (left). The gain that the AO system provides in both SNR (Figure 4 center) and integration time (Figure 4 right).

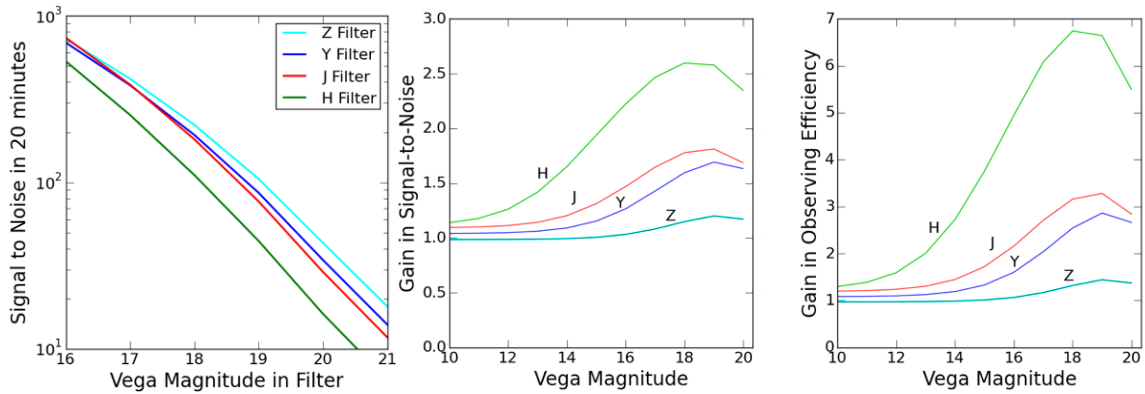

**Figure 4**. **Left:** The expected SNR and gains for the RTS system in the Z-band, Y-band, J-band, and H-band. The SNR is estimated by summing up the flux over the entire R=100 spectrum within each filter. The expected gain in the SNR **(center)** and efficiency **(right)** by including the adaptive optics system in RTS. The gain in efficiency is equivalent to the reduction in integration time needed to reach a specified SNR. We have assumed an integration time of 1200 s, square 0.8″ and 0.3″ apertures in seeing-limited and AO-corrected cases, respectively, and a Zenith angle of 30º.

The RTS spectrograph behind Robo-AO-2 will uniquely address the need for high-acuity and sensitive near-infrared spectral follow-up observations of tens of thousands of objects in mere months by combining an excellent observing site, unmatched robotic observational efficiency, and an AO system that significantly increases both sensitivity and spatial resolving power. We will initially use RTS to obtain the infrared spectra of ~4,000 Type Ia supernovae identified by the Asteroid Terrestrial-Impact Last Alert System over a two year period that will be crucial to precisely measuring distances and mapping the distribution of dark matter in the z < 0.1 universe, but more generally, this capability will be essential to many time domain classification projects.

**Organization, Partnerships, and Current Status:** Robo-AO was a partnership between California Institute of Technology and the Inter-University Centre for Astronomy and Astrophysics. Robo-AO KP was a partnership between the California Institute of Technology, the University of Hawai`i, the University of North Carolina at Chapel Hill, the Inter-University Centre for Astronomy and Astrophysics, and the National Central University, Taiwan. Robo-AO-2 is a partnership led by the University of Hawai`i, and including California Institute of Technology and the University of North Carolina at Chapel Hill. Robo-AO-2 is expected to see first light at the end of 2020. The Rapid Transient Surveyor spectrograph partnership additionally includes the University of California, San Diego.

**Cost Estimates:** Small <$20 M:
Robo-AO-2 system, <$2M;  Rapid Transient Surveyor Spectrograph, ~$2-3M

**References:**

Adams, E., et al., 2013, AJ, 146, 9.

Atkinson, D, et al., 2017, AJ, 153, 25.

Baranec, C., et al., 2013, JoVE, 72, e50021. http://www.jove.com/video/50021/

Baranec, C., et al., 2014, ApJ, 790, L8.

Baranec, C., et al., 2015, ApJ, 809, 70.

Baranec, C., et al., 2016a, AJ, 152, 18.

Baranec, C., et al., 2016b, SPIE 9909-0F. http://arxiv.org/abs/1606.07456

Baranec, C., Riddle, R., & Law, N., 2017, Handbook of Astronomical Instrumentation: Vol. 3. https://arxiv.org/abs/1709.07103

Baranec, C., et al., 2018, SPIE 10703-27. http://arxiv.org/abs/1806.01957

Bellm, E., et al., 2019, PASP, 131, 018002.

Chontos, A., et al., 2019, AJ, in press. https://arxiv.org/abs/1903.01591

Coughlin, M., et al., 2019, MNRAS, 485, 1412.

Furlan, E., et al., 2017, AJ, 153, 71 & 201.

Horch, E., et al., 2014, ApJ, 795, 60.

Jensen-Clem, R., et al., 2018, AJ, 155, 32.

Kraus, A., et al., 2016, AJ, 152, 8.

Kulkarni, S. R., 2012. IAU Symposium 285, 55-61.

Kurczynski, P., & Neff, J., 2018, SPIE 10706-03. https://arxiv.org/abs/1809.01294

Law, N., et al., 2014, ApJ, 791, 35.

Law, N., et al., 2016, SPIE 9907-0K. http://dx.doi.org/10.1117/12.2234446

Lillo-Box, J., et al., 2014, A&A, 566, A103.

Schonhut-Stasik, J., et al., 2017, AJ, 847, 97.

Schonhut-Stasik, J., et al., 2019, AJ, in press.

Severson, S., et al., 2013, SPIE 8617-09. https://arxiv.org/abs/1304.7280

Zeigler, C. et al., 2016, SPIE 9909-3Z. https://arxiv.org/abs/1608.00579

Ziegler, C., et al., 2017, AJ, 153, 66.

Ziegler, C., et al., 2018a, AJ, 155, 161.

Ziegler, C., et al., 2018b, AJ, 156, 83.

Ziegler, C., et al., 2018c, AJ, 156, 259.